# Self-organized plane arrays of metallic magnetic elements

M. Kostylev[(a)], R. Magaraggia[(a)], F.Y. Ogrin[(b)], V. Mescheryakov[(c)], N. Ross[(a)], and R.L. Stamps[(a)]

[(a)]School of Physics, University of Western Australia, Crawley 6009, WA, Australia
[(b)] School of Physics, University of Exeter, Exeter EX4 4QL, United Kingdom
[(c)]Moscow Institute of Radioelectronics and Automation (MIREA), Moscow, Russia

In the recent years much attention has been given to the study magnetic properties of periodically patterned metallic magnetic films. This area is important since magnetic periodical structures can be used for magnetic memory [1], magnetic logic [2], and microwave signal processing [3] applications.

Structuring centimetre-sized areas of films into dots or rings with diameters 100-400 nanometres using traditional lithography tools requires high precision instruments and is very time consuming. However there is another way to achieve the same goal in a more cost-efficient way. This is using a "natural lithography" based on a self-organization of polystyrene nanospheres on a hydrophilic surface into a highly in-plane periodic monolayer array. Such a structure formed on a surface of a multilayer stack containing magnetic metallic layers may serve as a lithographic mask for fabrication of a periodic array of magnetic dots [4]. Alternatively a nanosphere array may be deposited on a bare substrate and spacings between nanospheres may be filled with magnetic material using sputtering. This forms a honeycomb structure called "antidots" [5]. Figure 1 (taken from [4]) shows an AFM image of an exemplary array of nanodiscs fabricated using this technology. Such a perfect periodicity is observed over an area more than 1 cm in diameter.

In this work dynamic collective behaviour of the self-organized arrays was characterized by conventional and coplanar-waveguide [6] ferromagnetic resonance (FMR). A sample with the dot thickness of 25 nm, the mean dot diameter of 310 nm and the mean dot separation of 390 nm was studied. Fig. 2 shows results of measurements of in-plane and out-of-plane conventional FMR spectra for the sample. The out-of plane trace shows several peaks which may be attributed to different radial standing spin waves on the dots. The presence of several peaks shows that this fabrication technique is able to produce samples with small dispersion of geometrical and magnetic parameters. However the parameter dispersion is not entirely negligible, since one sees than the resonance line is about 10 times as large as expected for an unstructured Permalloy film. We explain this by a small difference in "magnetic" diameters and shapes of individual elements resulting in a small variation of resonance fields from dot to dot.

The frequency difference between the main in-plane and the main out-of-plane resonance is about 3.8 GHz. This value is much smaller than the usual one for unpatterned Permalloy films. This shows that the static internal magnetic field and static magnetization in the dots are considerably inhomogeneous. Furthermore, one should expect considerable effective dipole pinning of dynamic magnetization at the edges of the dots [8]. The work is now underway to extract the mean "magnetic" diameter of dots from these measurements.

Fig. 3 shows the results of the measurements using the coplanar-waveguide FMR technique. The measurements were done in a range of frequencies and magnetic fields applying the magnetic field in the plane of the array. Using this technique we were able to trace the ferromagnetic resonance response of the array in the range of frequencies 4-15 GHz. Note that this is not an easy task for this technique, since the amount of magnetic material contained in the sample is very small, and the measurements are done at the response levels close to the setup sensitivity threshold. The experimental point obtained by the conventional FMR technique is also shown in the figure. One sees very good agreement of results of both measurements.

ARC support under Discovery project "Magnetic nanostructures for emerging technologies" is gratefully acknowledged.

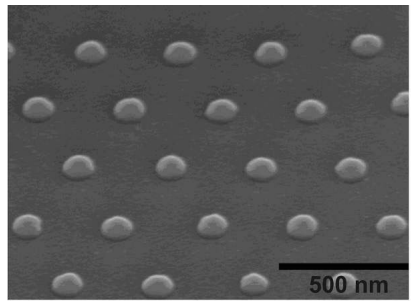

Fig. 1

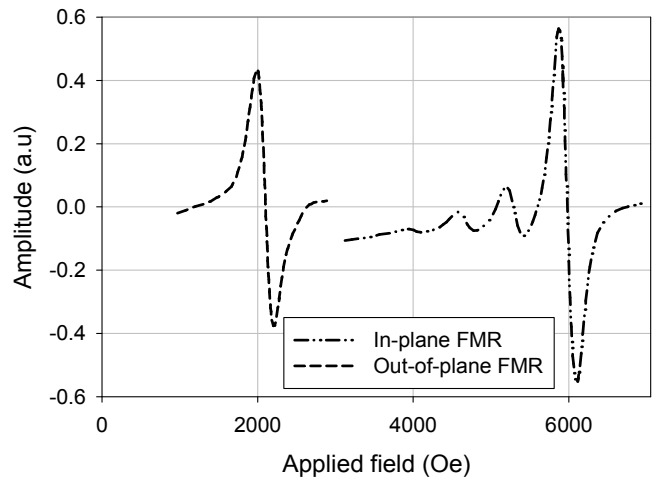

Fig. 2

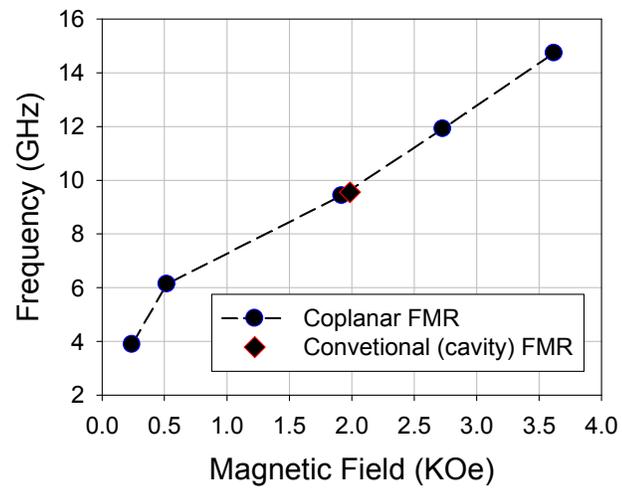

Fig. 3